\documentclass[epj]{svjour}
\usepackage{graphicx}
\newcommand{\be}{\begin{equation}}
\newcommand{\ee}{\end{equation}}
\newcommand{\bea}{\begin{eqnarray}}
\newcommand{\eea}{\end{eqnarray}}
\newcommand{\bann}{\begin{eqnarray*}}
\newcommand{\eann}{\end{eqnarray*}}
\newcommand{\bi}{\begin{itemize}}
\newcommand{\ei}{\end{itemize}}
\newcommand{\bcen}{\begin{center}}
\newcommand{\ecen}{\end{center}}

\begin{document}

\title{Neutral color-spin locking phase in neutron stars}
\author{Deborah N. Aguilera\inst{} 
\thanks{
supported by VESF-Fellowships EGO-DIR-112/2005.}
}                     
\offprints{deborah.aguilera@ua.es}          
\institute{Department of Applied Physics, University of Alicante, Apartado Correos 99, 03080 Alicante, Spain }
\date{Received: date / Revised version: date}
\abstract{We present results for the spin-1 color-spin locking phase (CSL) using 
a NJL-type model in two flavor quark matter for compact stars applications. 
The CSL condensate is flavor symmetric and therefore charge and color neutrality 
can easily be satisfied. We find small energy 
gaps $\simeq 1$ MeV, which make the CSL matter composition and the EoS 
not very different from the normal quark matter phase. We keep
finite quark masses in our calculations and obtain no gapless modes that could have 
strong consequences in the late cooling of neutron stars. 
Finally, we show that the region of the phase diagram relevant for neutron star cores,
when asymmetric flavor pairing is suppressed, could be covered by the CSL phase.	 
} 
\maketitle
\section{Introduction}
The investigation of cold dense quark matter
has received special attention 
due to the possible consequences for compact stars \cite{Alford:2002wf}. 
In particular, color super\-conduc\-ting quark matter phases enforcing color and charge neutrality 
has been widely studied \cite{Alford:2000sx}. 
Model calculations have shown that the intermediate density region of the neutral QCD phase diagram,
where the quark chemical potential is not sufficiently large to have the strange quark deconfined, 
might be dominated by $u,d$ quarks \cite{Ruster:2005jc}.  If this is the case, 
two-flavor quark matter phases may occupy a 
large volume in the core of compact stars \cite{Grigorian:2003vi}. 

On the other hand, local charge neutrality 
disfavor  the occurrence of phases with large gaps where quarks with 
different flavor pair in a spin-0 condensate, such as the 2SC phase \cite{Alford:2002kj}, 
in which quarks pair in e.g. $(u_rd_g)$ and $(d_ru_g)$ diquarks leaving the $u_b,d_b$ unpaired.  
Therefore, while the occurrence of 
neutral 2SC pure phase is rather model dependent and might be unlikely for
moderate coupling constants \cite{Aguilera:2004ag}, other  
phases such e.g. with spin-1 pairings \cite{Schafer:2000tw}, could be relevant 
for neutron star phenomenology. Specially, because these condensates with small energy 
gaps ($\Delta \simeq 1$ MeV) do not influence the equation of state (EoS) 
(it is not distinguishable from the normal quark (NQ) matter EoS) but 
they strongly affect the transport and thermal properties of quark matter \cite{Schmitt:2005wg}
and consequently the neutron star cooling. 
The unpaired quarks in the core lead to rapid cooling via the direct Urca process, 
incompatible with the observations. Phases that present no gapless modes prevent the direct 
Urca to work uncontrolled suppressing the neutrino emissivities 
and could explain the observed data \cite{Page:2000wt}. 

We consider in this work the color-antitriplet single-flavor spin-1 pairing in the Color-Spin-Locking 
(CSL) phase
\cite{Schafer:2000tw} and compare it with the 2SC phase. 
Our results for the CSL phase are obtained using the Nambu Jona-Lasinio (NJL) model \cite{Aguilera:2005tg} 
keeping finite quark masses and thus obtaining no gapless modes. 
Since these condensates are color neutral and single-flavor, neutral beta-equilibrated 
CSL quark matter is obtained easily. We present also the thermal behavior 
of the neutral CSL phase showing the phase diagram. 
Finally, we stress important features of the CSL phase that could give a 
consistent picture of a compact star with a superconducting quark matter core.

\section{Flavor symmetric (CSL) vs flavor asymmetric (2SC) pairing 
}

 We consider two flavor ($f=u,d$) quark matter, assuming that the 
 strange quark mass 
is large enough to appear only at higher densities.
In the NJL-type\footnote{We consider also 
 nonlocal extensions of the NJL model:  
the quark interactions act over a certain range 
introducing momentum dependent form factors in the 
the current-current interaction terms. 
The inclusion of high momenta states beyond the NJL-cutoff reduces  
the diquark condensates 
and lowers the density for the chiral phase transition (for a discussion 
see \cite{Aguilera:2006cj}).}  models
the quarks   interact locally by a 4-point vertex effective force. 
The NJL-Lagrangian 
$\mathcal{L}_{{\rm eff}}= \mathcal{L}_0 + \mathcal{L}_{q\bar q}+\mathcal{L}_{qq}$ 
contains a free part 
$\mathcal{L}_0$, 
 a quark-antiquark channel  
$\mathcal{L}_{q\bar q}$ that causes spontaneous chiral symmetry breaking with 
condensates
$\sigma = \langle \bar q q \rangle$ and a diquark channel 
$\mathcal{L}_{q q}$ that 
describes color superconductivity with condensate $\Delta$. 
The constituent quark mass is defined  
as $M=m-4G\sigma$ and the energy gaps $\Delta$ are listed in Tab.\ref{Tab_cond} for 
the phases: spin-0 2SC and spin-1 CSL. 

\begin{table}[hbt]
\caption{Two-flavor quark matter phases: flavor asymmetric spin-0 2SC and flavor symmetric spin-1 CSL. 
For $[S...A]$ the following pairs of indeces apply: $S,A= 3,2; 1,7;2,5$.}
\begin{flushleft}
\small{
\begin{tabular}{cccc}
phase&condensate $\Delta$ & diquarks & free \\
\hline
{\small 2SC}&
$2 G_1\langle~ \psi^{T}~C~\gamma_5~\tau_2~\lambda_2~\psi ~\rangle$&$u_rd_g,\,d_ru_g$
&$u_b,\,d_b$\\
(spin-0)&&&\\
\hline
{\small CSL}&
$4 H_v \langle~ \psi^{T}~C~\gamma_{[S}~\lambda_{A]}~\psi ~\rangle$&$u_ru_g,\,u_gu_b,$&-\\
(spin-1)&&$u_bu_r,\,(u\rightarrow d)$&
\end{tabular}
}
\end{flushleft}
\label{Tab_cond}
\end{table}
Linearizing $\mathcal{L}_{{\rm eff}}$ in the presence 
of the condensates
the thermodynamical potential $\Omega(T,\{\mu_f\})$
can be derived.
For the quark sector, in the 2SC case we obtain
\bea
\Omega_q&=&
4G\sigma^2+\frac{|\Delta|^2}{4G_1}
-2\!\! 
\sum_{\pm,j=1}^{3}\!\! 
\int
\!\!\left[
E_j^{\pm}
+2T \ln{(1+e^{-E_j^{\pm}/T})}
\right]\nonumber
\label{Omega2SC}
\eea
and for CSL, since the flavors decouple, $\Omega_q=\sum_f \Omega_f(T,\mu_f)$
\bea  
\Omega_f=
4G\sigma^2
+ \frac{3 |\Delta_f|^2}{8H_v}
-\sum_{k=1}^6 \int
\left[E^f_k+2T\ln{(1+e^{-E^f_k/T})}\right],
\nonumber
\label{omegaCSL} 
\eea  
 where $E_j^{\pm}$ and $E^f_k$ are the corresponding dispersion relations and the integration is in the momentum space. 

The lepton sector is modeled as an ideal electron gas with chemical potential $\mu_e$.  
 At the mean-field level, the
stationary points $\delta\Omega/\delta\Delta=\delta\Omega/\delta M=0$
define a set of gap equations for $\Delta$ and $M$. 
The stable solution is the one which corresponds to the absolute minimum of $\Omega$.

The parameters (quark mass $m=5$ MeV,  the coupling $G=4.66$ GeV$^{-2}$  
 and the cut-off  
$\Lambda=664$ MeV) 
are chosen to fit
the pion mass  
and the decay constant 
to a vacuum constituent 
quark mass equals $300$ MeV.  

\section{Results and Discussion}

\begin{figure}
\centering
{\includegraphics[width=0.8\columnwidth,angle=-90]{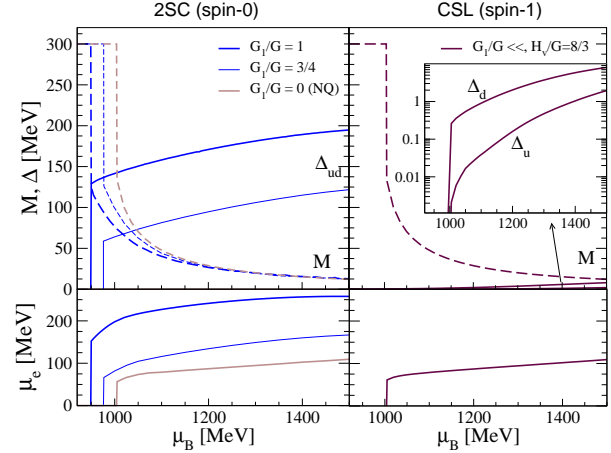}}
\caption{Gap equations solutions for neutral matter in the flavor asymmetric spin-0 2SC phase (on the left) and in the flavor symmetric spin-1 CSL phase (on the right) at $T=0$. The inset 
shows the small CSL gaps $\simeq 0.01-1 $ MeV.  
Upper panel: dynamical mass $M$ and energy gap $\Delta$ as a function of the baryon chemical potential $\mu_B$. 
Lower panel: Electron chemical potential $\mu_e$ vs $\mu_B$. Different 2SC couplings 
are considered from $G_1/G=1$ (dominant 2SC, see Fig.~\ref{fig:3}) 
down to $G_1/G=0$ ($\Delta \equiv 0$, NQ matter). The CSL coupling is taken as $H_v/G=8/3$. }
\label{fig:1}      
\end{figure}

We solve self-consistently the gap equations for the dynamical mass $M$ and the 
energy gap $\Delta$ for beta-equi\-li\-brated neutral matter: our solutions 
satisfy that the total quark electric charge $\mu_Q=-\mu_e$ and that the total 
charge density $n_Q-n_e=\frac{2}{3}n_u-\frac{1}{3}n_d-n_e$ vanishes. 
In Fig.~\ref{fig:1} we show $M$ and $\Delta$ as a function of the baryon 
chemical potential $\mu_B$
for the flavor asymmetric spin-0 2SC phase on the left and for the flavor 
symmetric spin-1 CSL phase on the right. 
While, for a fixed $\mu_B$, $M$ presents a similar magnitude for both phases, 
$\Delta$ differs by order of magnitudes: 
$\simeq 100$ MeV for 2SC, $\simeq 1$ MeV for CSL.  Moreover, 
the strength of the coupling constant determines whether the 2SC phase occurs: 
for {\it strong coupling} $G_1/G=1$, it presents large gaps $\Delta \simeq 150-200$ MeV 
that decrease as soon as the coupling does ($\Delta \simeq 50-100$ MeV 
for the usual Fierz value $G_1/G=3/4$) 
and vanishing for values lower than a critical one. 
The crucial point is that the coupling should be large enough to pair 
quarks of different flavors overcoming a large Fermi sea mismatch ($\simeq 60-80$ MeV). 
As a consequence, asymmetric flavor pairing might not be favorable unless the coupling 
is very strong (at least larger than 
$G_1/G=3/4$)\footnote{Actually, the critical value of $G_1/G$ at which the 2SC 
condensate breaks down is model and parameterization dependent. NJL calculations have 
obtained no pure 2SC at intermediate densities ($\mu_B=1200$ MeV) below $G_1/G=3/4$, 
see \cite{Ruster:2005jc,Aguilera:2004ag}. 
}. 
Therefore, flavor symmetric pairing becomes important when $G_1/G$ is not large enough
to have 2SC superconductivity. 

\begin{figure}
\centering
{\includegraphics[width=0.8\columnwidth,angle=-90]{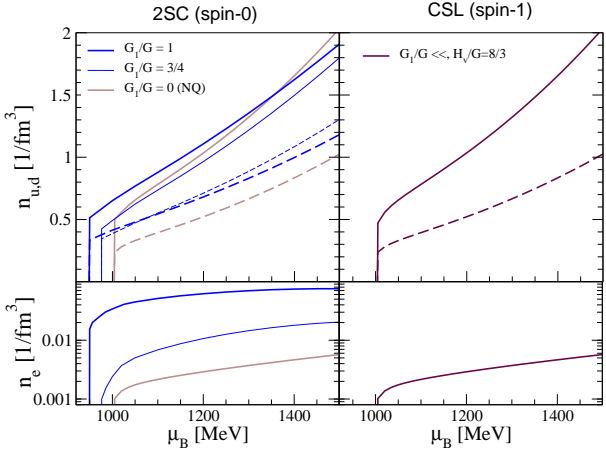}}
\caption{Quark ($d$ solid line, $u$ dashed line) and electron number densities for 2SC (on the left) and CSL (on the right) 
as a function of $\mu_B$ for neutral matter at $T=0$. 
}
\label{fig:2}      
\end{figure}

On the other hand, the occurrence of the flavor symmetric CSL pairing is 
not affected by the charge neutrality constraint. The CSL condensates, 
having small energy gaps in comparison to the free energy of the system, do not 
modify the thermodynamic properties respect to the normal phase. 
In Fig.~\ref{fig:2} we show the number density for quarks, $n_u,n_d$, and for electrons, $n_e$, as a function of $\mu_B$ for the 2SC (on the left) and for the CSL phase (on the right). Different values of the coupling for the 2SC phase from $G_1/G=1$ down to $G_1/G=0$ (NQ) are considered. 
We clearly see that $n_i^{2SC,G_1/G=0}=n_i^{NQ} \simeq n_i^{CSL}$ for 
each particle specie $i$, so the two phases, CSL and NQ, are indistinguishable 
from the matter composition. This conclusion holds also for other thermodynamic quantities
like pressure or energy density and therefore for the EoS. 

\begin{figure}
\centering
{\includegraphics[width=0.7\columnwidth,angle=-90]{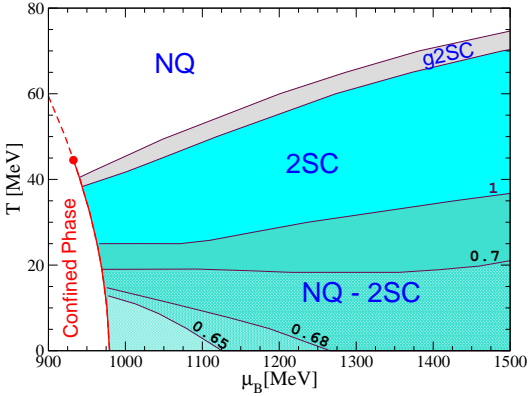}}
\caption{Phase diagram for the intermediate density region relevant for neutron 
star cores. For {\it strong coupling} 
($G_1/G\simeq1$), flavor asymmetric spin-0 2SC phase is dominant. The mixed phase 
NQ-2SC assures global charge neutrality. The volume fraction of the 2SC sub-phase 
is indicated by numbers over the corresponding lines 
(Gaussian form factor \cite{Aguilera:2004ag}). As $T$ increases, the volume fraction 
of the 2SC increases up to pure 2SC. Gapless 2SC is found before the transition 
to NQ occurs.  
}
\label{fig:3}      
\end{figure}

\begin{figure}
\centering
{\includegraphics[width=0.7\columnwidth,angle=-90]{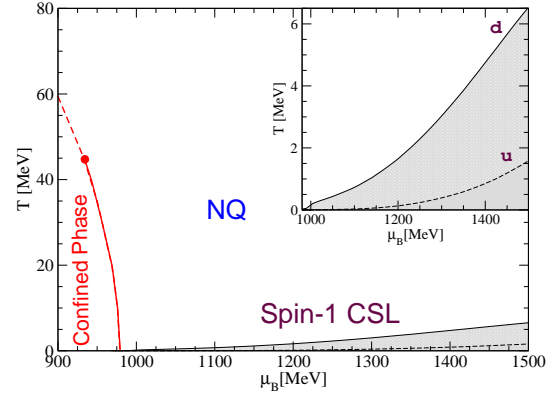}}
\caption{Idem as Fig.~\ref{fig:3} for 
{\it intermediate or weak coupling} ($G_1/G \leq 3/4$), 
for which flavor asymmetric pairing is no longer favorable. The volume fraction 
of the 2SC phase becomes very small ($< 10 \%$) and the structure of Fig.~\ref{fig:3} 
disappears. Thus, the matter could be in the normal state (NQ) or in a phase with 
flavor symmetric pairing such as the spin-1 CSL phase. }
\label{fig:4}       
\end{figure}

Finally, we found that while the phase diagram for {\it strong coupling} is 
dominated by the 2SC phase (Fig.~\ref{fig:3}) for {\it intermediate or weak coupling}, 
the CSL phase is favorable (Fig.~\ref{fig:4}).
Note that the low CSL critical temperatures $T_c^{CSL} \simeq 5$ MeV in contrast 
to the 2SC case, for which $T_c^{2SC} \simeq 50$ MeV.  Thus, we expect that in 
the late cooling of a neutron star, when the temperature has fallen below the MeV scale, 
a CSL superconducting quark core could develop. Stable hybrid star configurations 
have been obtained with a relatively large NQ matter core\cite{Grigorian:2003vi}, 
therefore, hybrid stars with a CSL superconducting core will be stable as well. 
Finally, a qualitative study of the interaction of the magnetic field with the 
CSL phase shows that a CSL-core is consistent with recent observations and 
models of magnetized neutron stars \cite{Aguilera:2006xv}. 

{\em Acknowledgments:}
D.N.A. thanks J.~A.~Pons, D.~Blaschke, N.~N.~Scoccola and J.~A.~Miralles for fruitful discussions.

\end{document}